\documentclass[9pt,conference]{IEEEtran}
\usepackage{dcase2026}
\usepackage[T1]{fontenc}
\usepackage[utf8]{inputenc}
\usepackage{amsmath,amssymb,graphicx,url,times,booktabs}

\title{Dual-Channel Contrastive Encoding of Sperm Whale Coda}
\name{João Victor Quintanilha$^{\dagger}$, \quad Emma Virnelli$^{\dagger}$, \quad Shuo Zhang
\thanks{$^{\dagger}$Equal contribution.}}
\address{Tufts University, Medford, MA, USA\\Joao.Quintanilha@tufts.edu, Emma.Virnelli@tufts.edu, Shuo.Zhang@tufts.edu}
\begin{document}
\maketitle
\begin{abstract}
Sperm whales (\emph{Physeter macrocephalus}) communicate through short click sequences called codas, whose structure includes the rhythmic pattern of inter-click intervals (ICI) and the spectral composition of the clicks. In this paper, we propose the Dual-Channel Contrastive Encoder (DCCE), which explicitly encodes rhythm and spectral information with separate encoders before fusing them. We evaluate DCCE on three coda-understanding tasks: social unit, coda type, and individual identification. We compare DCCE against strong baselines from literature under a common linear-probe protocol and analyze the role of each channel and training objective through ablations. On 1,383 codas, shared partitions and validation-only layer selection support three random splits and three recording-date-disjoint folds. DCCE with unit pairing and type supervision achieves mean random-split macro-F1 of 0.901/0.865/0.597 for unit/type/identity. Raw ICI remains strongest for type (0.987), and ICI plus temporal mel statistics is a strong identity reference (0.724). An auxiliary identity objective improves spectral-branch identity prediction but does not improve the mean fused-identity score in this configuration. Date-grouped evaluation and four year holdouts expose generalization limits. The results support task-dependent combinations of timing and spectral information, without establishing independent biological channels or general architectural superiority. Code, notebooks, and complete results are available at \url{https://dcce-research-site.vercel.app}.
\end{abstract}

\section{Introduction}
\label{sec:introduction}
Sperm whales (\emph{Physeter macrocephalus}) live in stable, matrilineal family units that belong to vocal clans, each clan sharing a distinct vocal dialect~\cite{andreas2022,gero2016}. That dialect is built from \emph{codas}: short, stereotyped sequences of clicks that whales exchange during social contact. Deciphering what codas encode is a central goal of computational bioacoustics and of the Cetacean Translation Initiative~\cite{andreas2022,paradise2025wham}. The \emph{rhythm} channel, the pattern of inter-click intervals (ICIs), contributes to coda \emph{type}: a category defined by click count and timing~\cite{gero2016,sharma2024}. The \emph{spectral} channel, the frequency content within each click, includes vowel- and diphthong-like patterns~\cite{begus2025,begus2026rspb}. The analogy to human voice motivates asking which features distinguish callers, without assuming that the two channels are independent.

The individual signal in a coda depends both on the coda type produced and on the set of whales one compares against. Gero et al.~\cite{gero2016} make these two dependencies concrete: discriminant analysis of \texttt{5R1} timing classified individuals within units J, U, and F at 73\%, 78\%, and 83\% accuracy, respectively, but at 46\% when individuals across units were pooled. The \texttt{1+1+3} type did not reliably discriminate individuals within units. Recovering \emph{who} produced a coda therefore depends on \emph{which} coda was produced and on how large and diverse the candidate pool is.

In this paper we propose the Dual-Channel Contrastive Encoder (DCCE), which learns a representation of sperm whale codas by encoding rhythm and spectral information with separate encoders followed by a fusion layer. Our guiding question is: \emph{does encoding rhythm and spectral texture as separate input channels, rather than learning implicitly from one input representation, improve what a model can learn from sperm whale codas?} The study contributes a dual-input architecture, matched supervision and channel ablations, and an evaluation of recording-context generalization.

\section{Related Work}
\label{sec:background}
\noindent\textbf{Coda structure and identity.} In the notation of Gero et al.~\cite{gero2016}, \texttt{5R1} denotes a five-click, regularly spaced type, while \texttt{1+1+3} groups five clicks into one, one, and three. Sharma et al.~\cite{sharma2024} showed that rhythm, tempo, rubato, and ornamentation yield at least 143 frequently realized combinations. Beguš et al.~\cite{begus2025} identified vowel- and diphthong-like spectral patterns; subsequent work reports interactions between vowel quality and conventional coda type~\cite{begus2026rspb}. Bermant et al.~\cite{bermant2019} already investigated neural sperm whale classification, including individual identity.

\noindent\textbf{Learned models for cetacean audio.} WhAM~\cite{paradise2025wham} provides a sperm-whale-specific reference, learning acoustic-token representations whose downstream classification ability is evaluated by probing. Broader bioacoustic references are AVES~\cite{hagiwara2023aves} and BirdNET~\cite{kahl2021birdnet}. Tolkova et al.~\cite{tolkova2026} applied BirdNET embeddings to individual identification using 234 upcalls of 11 North Atlantic right whales at three sites.

\noindent\textbf{Representation-learning lineages.} Self-supervised audio models such as wav2vec~2.0~\cite{baevski2020wav2vec} and HuBERT~\cite{hsu2021hubert} demonstrate learning speech representations from unlabeled audio. SimCLR uses the NT-Xent objective~\cite{chen2020simclr}, while CLIP learns aligned representations across image and text inputs~\cite{radford2021clip}. Finally, the speaker-verification lineage motivates mapping variable-length acoustic input to identity-relevant vectors: i-vectors and x-vectors provide statistical and neural approaches~\cite{dehak2011ivector,snyder2018xvector}, and metric-learning studies examine alternatives to classification losses~\cite{chung2020metric}. These precedents motivate the design, but neither establish its superiority on codas nor turn our closed-set ID task into open-set speaker verification.

\section{Methodology}
\label{sec:methodology}
\begin{figure}[t]
\centering\includegraphics[width=\columnwidth]{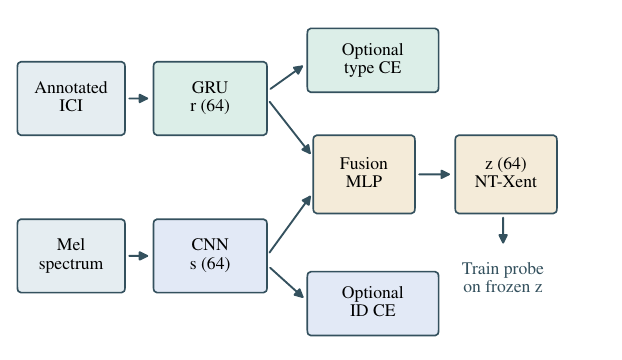}
\caption{DCCE: separate input encoders, fusion, and optional branch objectives. Positive-pair construction varies across conditions. The linear probe is trained after freezing the representation.}
\label{fig:architecture}
\end{figure}

We frame coda understanding as a representation-learning problem: given a coda, DCCE learns a single fixed-dimensional embedding from timing and spectral inputs, and we measure how much task-relevant structure that embedding captures with lightweight linear probes. Figure~\ref{fig:architecture} summarizes the design. DCCE is trained by combining a contrastive objective on $z$ with optional auxiliary classification losses on the branches.

\subsection{Rhythm Encoder}
The rhythm of a coda is captured by its sequence of inter-click intervals (ICIs): the elapsed times, in seconds, between consecutive clicks. A coda of $k$ clicks yields $k-1$ ICIs; since the codas in our data contain up to ten clicks, each coda has at most nine ICIs. Annotated ICIs are zero-padded on the right and standardized using training examples only. A two-layer GRU with hidden size 64 and inter-layer dropout 0.2 reads the sequence one ICI at a time; a linear projection and layer normalization map its final hidden state to $r\in\mathbb{R}^{64}$. We choose a compact recurrent encoder for these short sequences and the small dataset.

\subsection{Spectral Encoder}
The spectral channel is derived from a log-mel-spectrogram of the coda audio. Each recording is converted to mono and resampled to 48 kHz; native rates are 44.1, 48, 96, and 120 kHz. A short-time Fourier transform uses a 2,048-sample Hann window and a 1,152-sample hop, mapped onto 64 mel bands spanning 0--8 kHz. Log power is peak-relative, clipped to 80 dB, padded at the minimum level to a fixed $64\times256$ representation, and transformed by $x_{\mathrm{dB}}/40+1$. No mel input is truncated.

Three convolutional blocks (Conv2d--BatchNorm--ReLU--MaxPool, with 16, 32, and 64 channels) use $3\times3$ kernels and $2\times2$ pooling to reduce the input to a $64\times8\times32$ feature map. The flattened output passes through dropout 0.3 and a two-layer MLP with layer normalization to $s\in\mathbb{R}^{64}$. The full-coda spectrogram retains temporal structure, so this branch is not restricted to within-click identity cues.

\subsection{Fusion MLP}
We concatenate $r$ and $s$, then apply LayerNorm, Linear(128,64), ReLU, and Linear(64,64) to obtain $z$. Embeddings are normalized inside the contrastive loss. Primary probes use the saved fusion output; separate branch probes are post-hoc diagnostics.

\subsection{Training Objective}
The objective combines NT-Xent~\cite{chen2020simclr} with optional weighted cross-entropies for type on $r$ and identity on $s$. The type head encourages the rhythm branch to retain coda-type information rather than only unit-level structure. The ID head encourages individual discrimination, following the use of identity-classification supervision for speaker embeddings~\cite{snyder2018xvector}:
\begin{equation}
\mathcal{L}=\mathcal{L}_{\mathrm{NTXent}}+0.5a_T\mathcal{L}_{\mathrm{type}}+0.5a_I\mathcal{L}_{\mathrm{ID}},
\label{eq:objective}
\end{equation}
where $a_T,a_I\in\{0,1\}$. Inverse-frequency class weights are computed from training counts only; ineligible identity labels are masked. For positive views $i,j$, the contrastive term uses
\begin{equation}
\ell(i,j)=-\log\frac{\exp(\operatorname{sim}(z_i,z_j)/\tau)}{\sum_{k\ne i}\exp(\operatorname{sim}(z_i,z_k)/\tau)},\qquad\tau=0.07,
\end{equation}
with cosine similarity and a symmetric average over positive pairs. Assigning different auxiliary tasks to the branches encourages discriminative features but does not enforce statistical independence.

\subsection{Positive-Pair Construction}
The primary construction pairs distinct training codas $A,B$ from the same social unit: $f(r_A,s_A)$ and $f(r_B,s_B)$. This uses unit supervision. Actual channel swapping instead pairs $f(r_A,s_B)$ and $f(r_B,s_A)$. Without unit pairing, positives are independently augmented views of the same coda. Other batch views act as negatives, including possible same-unit examples; the objective is not a full supervised-contrastive treatment of all same-unit codas.

Each view masks a random four-bin frequency band and eight-frame time band, and adds Gaussian noise of SD 0.01 to standardized ICI entries corresponding to nonzero raw intervals. Padded intervals retain their standardized values. All main DCCE conditions use uniform shuffled sampling, 50 epochs of AdamW (learning rate $10^{-3}$, weight decay $10^{-4}$), cosine decay, and batch size 64. Temperature, auxiliary weights, sampling, and optimizer settings were fixed for this exploratory batch, not validation-tuned.

\subsection{DCCE Variants}
We denote unit pairing, type loss, and ID loss by U/T/I; three binary digits specify their presence. The main pair, DCCE 110 versus 111, differs only in ID loss. Both are task-informed models: absence of ID loss does not mean absence of supervision. We tested all eight U/T/I settings; selected contrasts are reported here, including 000, which removes task labels from the representation objective and sampler, but retains annotated ICI and supervised downstream probing.

Rhythm-only and spectrum-only controls apply both auxiliary heads to the available branch, with the same coefficients. Joint concatenation produces 128 dimensions without the fusion MLP; it is not independently trained late fusion. These architectures are evaluated with and without ID loss. Additional conditions test actual swapping and a 16-kHz mel ceiling; the latter also changes filter spacing at fixed band count.

\subsection{Linear Probe Evaluation}
All encoder weights are frozen after training. Balanced logistic regression predicts social unit, coda type, or individual ID. Features are standardized on training data, with fixed $C=1$ and at most 2,000 iterations. WhAM/AVES layers are selected per task using validation macro-F1 only, with earliest-layer tie breaking; BirdNET provides one representation. There is no training-plus-validation refit. Neural checkpoints use a fixed final epoch.

\section{Experiments}
\label{sec:experiments}
\begin{table*}[t]\centering\small
\caption{Main comparison: macro-F1 mean $\pm$ sample SD over three runs. Fixed/frozen rows train only the downstream probe locally; + ICI adds annotated timing. DCCE uses unit pairing (U), type loss (T), and optional ID loss (I), with uniform sampling. AVES FT uses unit/type/ID supervision while freezing the waveform frontend. All learned representations are frozen before probing. Primary scores exclude unsupported test classes; grouped ID coverage differs from random splits.}
\label{tab:main}
\begin{tabular}{lcccccc}\toprule
&\multicolumn{3}{c}{Random split}&\multicolumn{3}{c}{Recording-date grouped}\\
Model & Unit & Type & ID & Unit & Type & ID\\\midrule
Raw ICI & $0.570\pm0.023$ & $0.987\pm0.009$ & $0.439\pm0.007$ & $0.540\pm0.036$ & $0.956\pm0.036$ & $0.379\pm0.103$\\
Temporal mel & $0.773\pm0.007$ & $0.202\pm0.025$ & $0.429\pm0.045$ & $0.569\pm0.094$ & $0.069\pm0.012$ & $0.130\pm0.074$\\
ICI + temporal mel & $0.837\pm0.013$ & $0.804\pm0.016$ & $0.724\pm0.039$ & $0.608\pm0.150$ & $0.668\pm0.051$ & $0.255\pm0.062$\\
\midrule
WhAM (frozen) & $0.906\pm0.011$ & $0.258\pm0.046$ & $0.450\pm0.052$ & $0.667\pm0.155$ & $0.066\pm0.016$ & $0.101\pm0.075$\\
AVES (frozen) & $0.910\pm0.010$ & $0.253\pm0.028$ & $0.491\pm0.029$ & $0.700\pm0.100$ & $0.061\pm0.022$ & $0.074\pm0.051$\\
BirdNET (frozen) & $0.887\pm0.020$ & $0.215\pm0.053$ & $0.480\pm0.022$ & $0.612\pm0.133$ & $0.063\pm0.011$ & $0.085\pm0.022$\\
\midrule
WhAM + ICI (frozen) & $0.921\pm0.005$ & $0.722\pm0.027$ & $0.598\pm0.028$ & $0.768\pm0.057$ & $0.376\pm0.151$ & $0.153\pm0.135$\\
AVES + ICI (frozen) & $0.926\pm0.011$ & $0.667\pm0.044$ & $0.652\pm0.044$ & $0.712\pm0.111$ & $0.366\pm0.084$ & $0.128\pm0.103$\\
BirdNET + ICI (frozen) & $0.884\pm0.018$ & $0.497\pm0.028$ & $0.580\pm0.031$ & $0.628\pm0.095$ & $0.287\pm0.061$ & $0.100\pm0.055$\\
\midrule
DCCE U+T & $0.901\pm0.020$ & $0.865\pm0.115$ & $0.597\pm0.051$ & $0.598\pm0.078$ & $0.692\pm0.088$ & $0.248\pm0.130$\\
DCCE U+T+I & $0.907\pm0.010$ & $0.735\pm0.091$ & $0.581\pm0.012$ & $0.596\pm0.052$ & $0.628\pm0.074$ & $0.223\pm0.175$\\
\midrule
AVES FT & $0.924\pm0.007$ & $0.243\pm0.025$ & $0.414\pm0.078$ & $0.626\pm0.157$ & $0.072\pm0.050$ & $0.060\pm0.026$\\
AVES + ICI FT & $0.913\pm0.011$ & $0.466\pm0.059$ & $0.544\pm0.017$ & $0.641\pm0.158$ & $0.211\pm0.069$ & $0.073\pm0.038$\\
\bottomrule\end{tabular}\end{table*}

\begin{figure*}[t]
\centering\includegraphics[width=\textwidth]{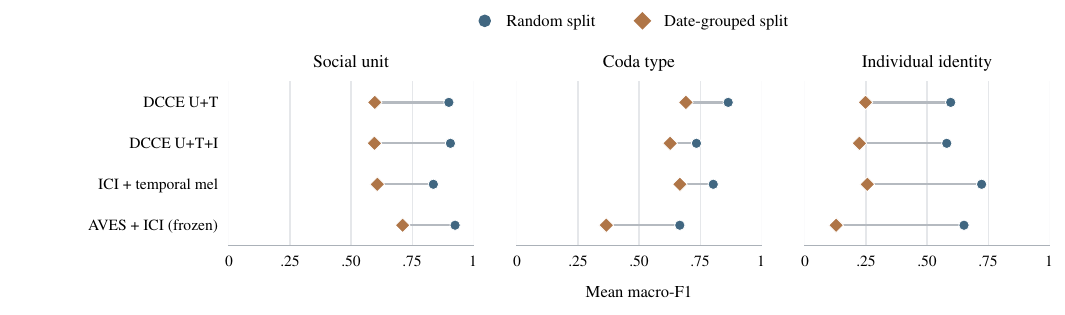}
\caption{Random versus recording-date-grouped evaluation for four representative models. Points show mean macro-F1 over three runs; sample SD and the full model comparison appear in Table~\ref{tab:main}. Lines connect protocol means. Grouped identity evaluation covers different supported classes and examples, so these differences are not isolated estimates of a recording-context penalty.}
\label{fig:generalization}
\end{figure*}
\subsection{Dataset and Evaluation Protocol}
The dataset used in our experiment consists of 1,501 audio recordings from the Dominica Sperm Whale Project (DSWP) public HuggingFace release (\texttt{orrp/DSWP})~\cite{paradise2025wham}. After excluding 118 noise-flagged files, 1,383 clean codas remain, spanning three social units (A, D, F) and 22 coda types. Labels (social unit, coda type, individual ID) are reconstructed from \texttt{DominicaCodas.csv}, using its \texttt{codaNUM2018} index to match DSWP file indices. Reconstructed metadata place these codas in 2005--2010 and clan EC1. ID evaluation uses 754 codas from 11 individuals: 620 unknown-ID codas, eight ambiguous two-individual annotations, and one identified singleton are excluded from ID training/probing, while remaining eligible for other tasks. Severe class imbalance (unit F contributes 821/1,383 codas, or 59.4\%) motivates macro-F1 and balanced class weights. All tasks share the same coda partition, keeping every locally trained representation separate from its test examples.

We use three random splits (seeds 0--2), approximately 64/16/20\% train/validation/test, stratifying eligible identities and otherwise unit. Recording-date splits use the first three outer folds (indices 0--2) of a five-fold stratified-group partition (shuffle seed 0), with grouped validation partitions. Dates are session proxies. Four additional year holdouts exclude the test year completely; validation uses a date-grouped fold from the remaining years, and neural seed is fixed at 0.

Primary macro-F1 averages over true test classes observed in training, on supported examples. Grouped ID support is 186/213, 111/121, and 145/145 eligible examples, spanning three, six, and six scored identities. Thus random/grouped ID scores describe different supported populations. Full-eligible secondary scores include unseen classes. We report means and sample SD over three runs, not confidence intervals.

The design is exploratory: earlier unit-balanced experiments informed these comparisons, and year/input-matched controls followed partial main results. Earlier training-regime scores are not pooled with the matched uniform-sampling ablations.

\subsection{DCCE vs. Baselines}
We evaluate three literature-grounded baselines under the same splits and linear-probe protocol as DCCE: WhAM, AVES, and BirdNET. Fixed references are raw ICI, temporal mel means/std within four valid-frame bins (512 dimensions), and their concatenation (521 dimensions). Temporal binning retains coarse timing that a single time-averaged mel vector would discard. Frozen WhAM, AVES, and BirdNET are evaluated both alone and concatenated with ICI. These last controls match timing-input availability, with only probes trained locally.

Pretrained frontends retain native procedures: AVES at 16 kHz over the full segment; the released WhAM codec at 44.1 kHz with mono/loudness processing; BirdNET at 48 kHz with a three-second window. Three waveforms exceed that window.

Supervised AVES adaptation updates the projection and all 12 transformer blocks, keeping the waveform CNN frozen. Unit/type/ID loss weights are 1/0.5/0.5, with uniform sampling. Ten fixed AdamW epochs use backbone/head rates $2\times10^{-5}/10^{-3}$, weight decay $10^{-4}$, microbatches of four accumulated to 64, and clip gradients at~1. Labels are comparable in availability, but objectives, capacity, pretraining, and budgets remain different.

\subsection{Results}
No single model wins on all three tasks (Table~\ref{tab:main}). Raw ICI attains $0.987\pm0.009$ random type macro-F1; ICI plus temporal mel attains $0.724\pm0.039$ ID macro-F1. Including unseen type labels changes the raw-ICI random/grouped type means from 0.987/0.956 to 0.968/0.919. Strong timing references are essential when type labels themselves reflect interval structure.

DCCE 110 attains random unit/type/ID means of 0.901/0.865/0.597. Its ID mean exceeds either separately probed branch (rhythm 0.459, spectrum 0.367), but remains below the simple combined-feature reference. DCCE 111 gives 0.907/0.735/0.581: adding ID loss does not improve mean fused-ID performance in this configuration. These descriptive three-run differences do not establish a universal effect of identity supervision.

Frozen AVES plus ICI attains random ID 0.652, versus 0.491 for audio alone. Fine-tuned AVES plus ICI attains 0.544, versus 0.414 without ICI. Input availability therefore materially changes the comparison, while these fixed adaptation schedules do not establish optimal pretrained performance.

\subsection{Ablation Study}
\begin{table}[t]\centering\small
\setlength{\tabcolsep}{3pt}
\caption{Matched architecture controls: macro-F1 mean $\pm$ sample SD, three runs. R/G denote random/date-grouped; digits denote U/T/I. All use uniform sampling, 50 epochs, and whole-coda pairing. Enabled auxiliary heads act on the available branch; representations are frozen before probing. Joint concatenation is 128-dimensional; other outputs are 64-dimensional. Fused DCCE anchors appear in Table~\ref{tab:main}.}
\label{tab:core_ablations}
\begin{tabular}{lccc}\toprule
Model & R Type & R ID & G ID\\\midrule
Rhythm 110 & $.978\pm.021$ & $.477\pm.037$ & $.352\pm.134$\\
Rhythm 111 & $.973\pm.013$ & $.454\pm.061$ & $.331\pm.143$\\
\midrule
Spectrum 110 & $.202\pm.021$ & $.358\pm.103$ & $.054\pm.047$\\
Spectrum 111 & $.199\pm.037$ & $.444\pm.092$ & $.092\pm.016$\\
\midrule
Joint concat. 110 & $.939\pm.062$ & $.636\pm.015$ & $.275\pm.117$\\
Joint concat. 111 & $.927\pm.060$ & $.696\pm.055$ & $.239\pm.043$\\
\bottomrule\end{tabular}\end{table}

Without auxiliary losses, same-unit pairing changes random unit/ID means from 0.693/0.624 (000) to 0.861/0.434 (100). The grouped ID ordering reverses, 0.181 versus 0.245. Pairing therefore imposes task- and partition-dependent tradeoffs.

Table~\ref{tab:core_ablations} compares channel and concatenation controls under matched label access, against the fused DCCE rows in Table~\ref{tab:main}. Neither these results nor the dimensionality-different concatenation control establish an exclusively spectral identity code or a necessary benefit from the fusion MLP.

Actual swapping raises random ID from 0.581 to 0.657 relative to whole-coda 111, while grouped ID falls from 0.223 to 0.189. A 16-kHz ceiling raises random ID to 0.631 without and 0.623 with ID loss; grouped scores are 0.250 and 0.173. These are bandwidth sensitivities, not localization of biological identity cues.

Post-hoc probes of fixed checkpoints expose a branch/fusion distinction: ID supervision raises random spectral-branch ID from 0.367 to 0.443, while fused ID falls from 0.597 to 0.581. Rhythm-branch type remains 0.983/0.982 without/with ID loss, whereas fused type is 0.865/0.735. Auxiliary branch performance need not carry through fusion.

\subsection{Recording Context and Generalization}
Explicit day/month/year parsing includes all 1,383 codas. Year-by-unit counts (A/D/F) are 10/0/358 in 2005, 27/171/65 in 2008, 30/18/47 in 2009, and 174/132/351 in 2010. Pearson $\chi^2=530.04$ (df=6), with uncorrected Cramér's $V=\sqrt{\chi^2/[N\min(3,2)]}=0.438$. This is an association in the dataset, not proof of a model's causal reliance on recording artifacts.

Figure~\ref{fig:generalization} shows lower grouped means across all three tasks for the four illustrated models, with task-dependent orderings. Holding out 2005/2008/2009/2010 yields unit macro-F1 of 0.592/0.362/0.767/0.346 for DCCE 110 and 0.448/0.391/0.713/0.357 for DCCE 111. Each year uses one fit; 2005 contains two test units, other years three. Both variants score below frozen AVES in every year; these outcomes do not establish general protection against recording-context variation.

We report selected unit/type ID diagnostics. For the previously discussed \texttt{1+1+3} type, DCCE 110's grouped-fold ID macro-F1 is 0.000/0.233/0.234 (52/53/34 codas; 3/3/4 identities); single-identity strata are unscored. Separate within-unit probes require at least two supported identities in train and test. Only unit F supports these probes on all grouped folds: mean ID is 0.384 for raw ICI, 0.267 for DCCE 110, and 0.238 for DCCE 111. These comparisons reduce one group-membership shortcut without removing all type/context dependence.

\section{Discussion}
\subsection{Interpretation and Implications}
DCCE's performance across the three tasks supports studying rhythmic and spectral information as separate inputs, but does not assign each biological attribute exclusively to one branch. Fusion can improve over individually probed branches, while simple combined features remain strong references. For small bioacoustic datasets, these findings motivate matching supervision and annotation access, retaining simple temporal controls, and evaluating across recording groups. The branch/fusion distinction makes cross-channel coupling a design question for future work rather than a fixed architectural choice. Objective interactions or compression are possible explanations, not mechanisms established by these experiments.

\subsection{Limitations}
Following our analysis, a key limitation is the dataset itself: a single Dominica population with three social units, all from the EC1 clan. Another restriction is that there are no vowel annotations for the 1,383 codas evaluated here: the available hand-annotated vowel metadata covers different coda identifiers. The spectral branch is trained through the fused contrastive objective and optional ID classification, not explicit phonological categories.

Annotated ICI is additional information unavailable to a waveform-only system. Closed-set ID tests do not establish recognition of unseen animals or communicative meaning. Sparse identity coverage and three runs limit precision; recording dates are not verified sessions.

WhAM's published training uses DSWP+CETI~\cite{paradise2025wham}, and this evaluation also uses DSWP. Exact checkpoint training membership is unavailable, so local holdouts do not establish external pretraining independence. The exploratory design and heterogeneous model histories further limit causal or universal comparisons.

\section{Conclusion and Future Work}
We asked whether encoding rhythm and spectral texture separately before fusion improves what a model can learn from sperm whale codas. DCCE combines useful information across unit, type, and identity tasks, but gains depend on supervision and evaluation conditions. These experiments do not establish a benefit of learned fusion over joint concatenation. A further question is whether DCCE generalizes to whale populations beyond the single Dominica clan studied in this work. Beyond adjusting the dataset, future work could also treat the strength of the identity objective as a tunable parameter, testing how auxiliary weighting and fusion affect the retention of type and identity information. Such evaluations should use independent recordings and explicit annotation budgets. The paper, code, executed notebooks, complete result tables, and audio examples are available at \url{https://dcce-research-site.vercel.app}.

\clearpage
\bibliographystyle{IEEEtran}
\bibliography{refs}
\end{document}